\documentclass[pt 12, onecolumn,oneside, reprint]{pnas-new}
\usepackage{graphicx,amsmath}
\usepackage[flushleft]{threeparttable}
\usepackage{siunitx}
\usepackage{setspace}
\setboolean{displaywatermark}{false}

\usepackage{amssymb,amsfonts}

\usepackage{bm}
\usepackage{color}
\usepackage{upgreek}
\usepackage{epstopdf}
\usepackage{lineno}

\usepackage{multirow}

\templatetype{pnasresearcharticle} % Choose template
\title{Giant and explosive plasmonic bubbles by delayed nucleation}

\author[a, b, 1, *]{Yuliang Wang}
\author[b, c, 1]{Mikhail E. Zaytsev}
\author[b, d]{Guillaume Lajoinie}
\author[d, e]{Hai Le The}
\author[d, e]{Jan C. T. Eijkel}
\author[d, e]{Albert van den Berg}
\author[b, d]{Michel Versluis}
\author[f]{Bert M. Weckhuysen}
\author[g, b]{Xuehua Zhang}
\author[c]{Harold J. W. Zandvliet}
\author[b, h, *]{Detlef Lohse}

\affil[a]{Robotics Institute, School of Mechanical Engineering and Automation, Beihang University, Beijing 100191, P.R. China}
\affil[b]{Physics of Fluids, Max Planck Center Twente for Complex Fluid Dynamics and J.M. Burgers Centre for Fluid Mechanics, MESA+ Institute for Nanotechnology, University of Twente, P.O. Box 217, 7500AE Enschede, The Netherlands}	
\affil[c]{Physics of Interfaces and Nanomaterials, MESA+ Institute for Nanotechnology, University of Twente, 7500AE Enschede, The Netherlands}
\affil[d]{MIRA Institute for Biomedical Technology and Technical Medicine, University of Twente, Enschede, The Netherlands}
\affil[e]{BIOS Lab-on-a-chip, MESA+ Institute for Nanotechnology, University of Twente, University of Twente, P.O. Box 217, 7500AE Enschede, The Netherlands}
\affil[f]{Inorganic Chemistry and Catalysis, Debye Institute for Nanomaterials Science, Utrecht University, Universiteitsweg 99, 3584 CG Utrecht, The Netherlands}
\affil[g]{Department of Chemical and Materials Engineering, University of Alberta, 12-211 Donadeo Innovation Centre for Engineering, Edmonton, Alberta, Canada}
\affil[h]{Max Planck Institute for Dynamics and Self-Organization, Am Fassberg 17, 37077 G\"ottingen, Germany}

\leadauthor{Wang}

\significancestatement{Plasmonic microbubbles are at the center of numerous applications, including micro/nanomanipulation, biomedical diagnosis and therapy, and solar energy harvesting. Controlling the dynamics of these microbubbles is key to successfully exploit their potential in the growing number of applications. Literature has vastly overlooked the nucleation and early microsecond dynamics by which these microbubbles arise. It is of utmost importance to understand this early phase, as it displays the most violent dynamics and is the precursor and origin of all subsequent dynamics, on which the many applications rely.}

\authorcontributions{Y.W., M.E.Z, X.Z., H.J.W.Z. and D.L. designed research; Y.W., M.E.Z and G.L., performed the experimental research; G.L. and D.L. performed the theoretical research, H.L.T., J.C.T.E., A.vdB., M.V., B.M.W., and X.Z. contributed new reagents tools; Y.W., M.E.Z and G.L. analyzed data; and Y.W., M.E.Z, G.L., H.J.W.Z. and D.L. wrote the paper.}
\authordeclaration{The authors declare no conflict of interest.}
\equalauthors{\textsuperscript{1}Y.W.(Yuliang Wang) and M.Z. (Mikhail E. Zaytsev) contributed equally to this work.}
\correspondingauthor{\textsuperscript{*} To whom correspondence should be addressed. E-mail: d.lohse$@$utwente.nl and wangyuliang$@$buaa.edu.cn}


\begin{abstract}
When illuminated by a laser, plasmonic nanoparticles immersed in water can very quickly and strongly heat up, leading to the nucleation of so-called plasmonic vapor bubbles, which have huge application potential in  e.g.\ solar light-harvesting, catalysis, and for medical applications. Whilst the long-time behavior of such bubbles has been well-studied, here, by employing ultra-high-speed imaging, we reveal the nucleation and early life phase of these bubbles. After some delay time $\tau_d$ after beginning of the illumination, a giant bubble explosively grows, up to a maximal radius of 80 $\mu m$, and collapses again within $\approx200 \mu s$ (bubble life phase 1). The maximal bubble volume $V_{max}$ remarkably increases with decreasing laser power $P_\ell$. To explain this behavior, we measure the delay time $\tau_d$ from the beginning of the illumination up to nucleation, which drastically increases with decreasing laser power, leading to less total dumped energy $ E = P_\ell \tau_d$. This  dumped energy $E$ shows a universal linear scaling relation with $V_{max}$, irrespectively of the gas concentration of the surrounding water. This finding supports that the initial giant bubble is a pure vapor bubble. In contrast, the delay time does depend on the gas concentration of the water, as gas pockets in the water facilitate an earlier vapor bubble nucleation, which leads to smaller delay times and lower bubble nucleation temperatures. After the collapse of the initial giant bubbles, first much smaller oscillating bubbles form out of the remaining gas nuclei (bubble life phase 2, up to typically $10 ms$). Subsequently the known vaporization dominated growth phase takes over and the bubble stabilizes (life phase 3). In the final life phase 4 the bubble slowly grows by gas expelling due to heating of the surrounding. Our findings on the explosive growth and collapse during the  early life phase of a plasmonic vapor bubble have strong bearings on possible applications of such bubbles, affecting their risk assessment.
\end{abstract}

\dates{This manuscript was compiled on \today}
\begin{document}

% Optional adjustment to line up main text (after abstract) of first page with line numbers, when using both lineno and twocolumn options.
% You should only change this length when you've finalised the article contents.
%\verticaladjustment{-2pt}

\maketitle
\thispagestyle{firststyle}
\ifthenelse{\boolean{shortarticle}}{\ifthenelse{\boolean{singlecolumn}}{\abscontentformatted}{\abscontent}}{}

% If your first paragraph (i.e. with the \dropcap) contains a list environment (quote, quotation, theorem, definition, enumerate, itemize...), the line after the list may have some extra indentation. If this is the case, add \parshape=0 to the end of the list environment.

\dropcap{N}oble metal nanoparticles under resonant irradiation of continuous wave (cw) lasers can produce a huge amount of heat due to the excitation of their plasmon resonance frequency, resulting in the vaporization of the surrounding water and the formation of micro-sized plasmonic bubbles \cite{neumann2013, baffou2014, carlson2012, fang2013, baral2014,  liu2015}.
%Continuous wave (cw) laser irradiation has been shown to induce the formation of microbubbles around water-immersed plasmonic nanoparticles \cite{neumann2013, baffou2014, carlson2012, fang2013, baral2014,  liu2015}.
These bubbles appear in numerous applications, including micro/nanomanipulation \cite{zhao2014, lin2016}, biomedical therapy\cite{lapotko2009, emelianov2009, baffou2013, shao2015, liu2014}, and solar energy harvesting \cite{neumann2013, neumann2013-1, fang2013, boriskina2013, ghasemi2014, baral2014, guo2017}. Understanding the nucleation mechanism and growth dynamics of these plasmonic bubbles is key to successfully take up the challenges connected to these applications. However, most studies up to now have not yet focused on the plasmonic microbubble {\it nucleation} and its early dynamics but, instead, are conducted on the long-term (milliseconds to seconds) timescale \cite{baffou2014jpc, liu2015, chen2017, baral2014}. In a recent study, we revealed that the long-time growth of these plasmonic bubbles can be divided into two phases, namely a vaporization dominated phase in which vapor bubbles grow (up to 10 ms), followed by a slow diffusion-dominated growth \cite{wang2017}. This later phase reflects the role of dissolved gas in the growth dynamics of the bubbles \cite{wang2017}. Note that the vaporization event for plasmonic bubbles is different than for normal vapor bubbles, which arise from simply locally heating the liquid with a laser \cite{sun2009, tagawa2012, zwaan2007, quinto-su2008, prosperetti2017}.

Upon laser irradiation, water around plasmonic nanoparticles at solid-liquid interfaces experiences a rapid temperature increase, first proportional to the input laser intensity. The resulting temperature rise can exceed the boiling temperature (100$^{\circ}$C) within a few nanoseconds to microseconds  \cite{hleb2010, carlson2012, lombard2014, hou2015, lombard2017}. This is several orders of magnitude faster than the millisecond time scale in which plasmonic bubbles are normally observed. At the $n$s to $\upmu$s time scales, the fate of the plasmonic nanoparticles under cw laser irradiation and that of the liquid in their vicinity has remained unexplored. The reason for this primarily lies in the difficulty to visualize the early stage of the vaporization dynamics around the plasmonic nanoparticles, due to the lack of imaging systems with sufficient temporal resolution.

In this paper, we overcome this bottleneck by means of the ultra-high-speed imaging facility Brandaris128 \cite{chien2003, gelderblom2012} and reveal the early dynamics of plasmonic bubbles nucleating on an immersed gold nanoparticle (GNP) decorated surface. Brandaris128 has a temporal resolution of 100 ns, which allowed us to reveal that a giant transient vapor bubble arises prior to the hitherto observed plasmonic bubbles (Supplemental material, video 1). The delay between the beginning of the laser heating and the bubble nucleation depends on the laser power and the concentration of the gas dissolved in water. We compare a gas-rich and a gas-poor case with the latter having about half of the gas concentration as the former. The measured relation between the delay and the laser power will be used to estimate the nucleation temperature of the vapor bubbles in water. Combined with the sub-microsecond cavitation dynamics, this nucleation delay provides information on the energy conversion efficiency. Our findings have strong bearings on the applications mentioned above, and affect their risk assessment.

\begin{figure*}[t]
	\centering
	%	\begin{minipage}[c]{0.75\textwidth}
	\includegraphics[width=0.8\textwidth]{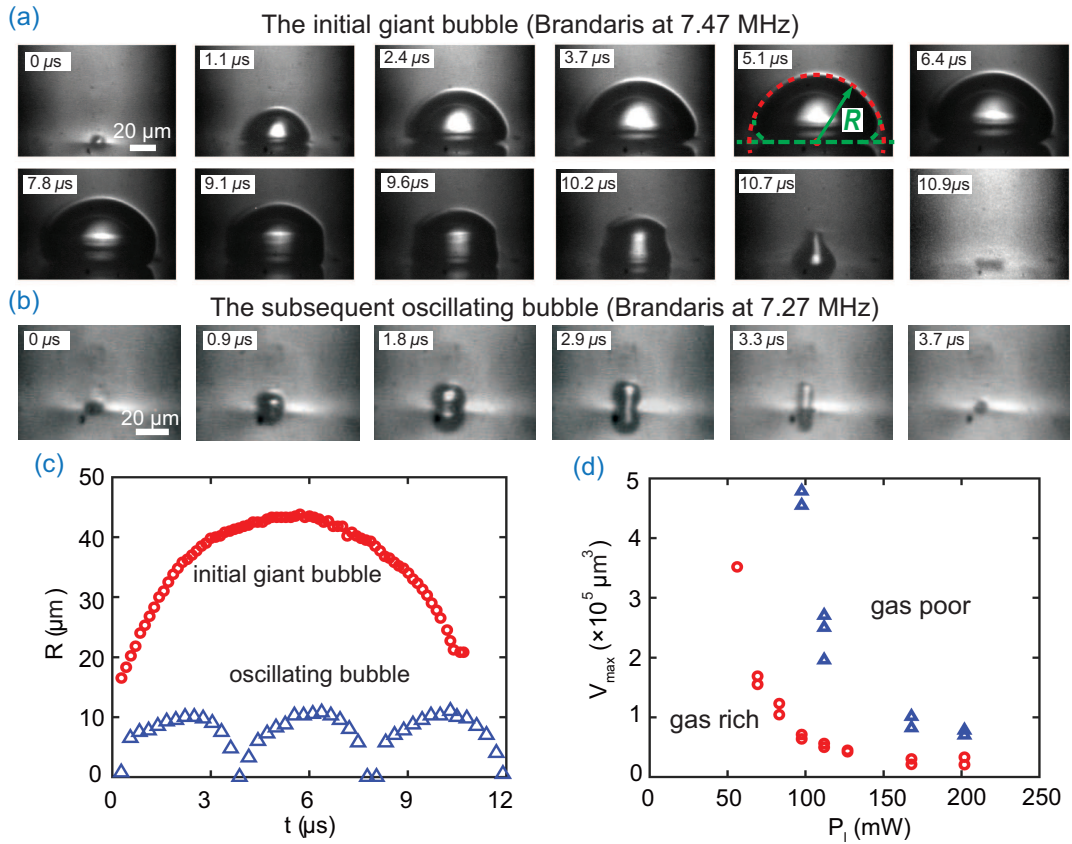}
	%	\end{minipage}\hfill
	%	\begin{minipage}[c]{0.25\textwidth}
	\caption{	
		Evolution of an initial giant bubble (a) and a subsequent oscillating bubble (b) along with their life cycles captured at 7.47 Mfps for the gas-rich case and $P_l=185mW$. The two kinds of bubbles show different shapes and dynamics. (c) Radius of curvature $R$ as a function of time for an initial giant bubble and for the subsequent oscillating bubbles.  (d) Maximum volume $V_{max}$ of the giant bubble as function of laser power $P_l$ in gas-rich water and gas-poor water. Counter-intuitively, the bubble volume decrease with increasing laser power $P_l$.
	}
	\label{fig:huge_vs_small}
	%	\end{minipage}
\end{figure*}

An array of GNPs with diameters of 100 nm and spaced by 260 nm (center to center) was deposited on a fused silica substrate to induce plasmonic bubble nucleation. Experiments were first conducted with the Brandaris 128 ultra-high-speed imaging system at frame rates near 8 Mfps. The origin of time ($t=0$s) is the instant at which the laser beam hits the substrate. After a delay time $\tau_d$, a bubble (Fig. \ref{fig:huge_vs_small}a) nucleates. During the violent subsequent growth, the cavitation bubble can reach a size exceeding 100 $\upmu$m within 6 $\upmu$s, while retaining a hemi-spherical shape. Fig. \ref{fig:huge_vs_small}d shows that an increase of the laser power $P_l$ counter-intuitively leads to a decrease in the maximum bubble volume $V_{max}$. Also counter-intuitively, the gas-poor case leads to larger bubbles as compared to the gas-rich case.

After the initial giant bubble has collapsed, smaller bubbles experience cycles of sustained oscillations (Fig. \ref{fig:huge_vs_small}b and Supplemental material video 2) before gradually stabilizing. These bubbles, referred to as oscillating bubbles, are about 100$\times$ smaller in volume than the initial giant bubble (Fig. \ref{fig:huge_vs_small}c). Accordingly, the oscillation period of these bubbles is substantially shorter than the lifetime $\tau_c$ of the giant bubble.

\begin{figure*}[t]
	\vspace{-10pt}
	\centering
	\includegraphics[width=1.0\textwidth]{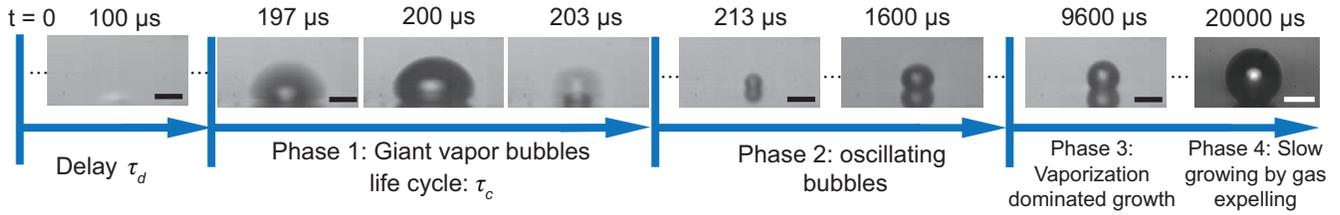}
	\caption{
		Time sequence of the bubble dynamics under continuous laser irradiation on the patterned GNP sample surface in gas-rich water and $P_l$ = 83 mW. According to their nucleation and growth dynamics, the evolution of the plasmonic bubbles is divided into four phases. The scale bar is 25 $\upmu$m.
	}
	\label{fig:table}
	\vspace{-10pt}
\end{figure*}

These two life phases, {\it i.e.} the giant bubble growth and collapse (life phase 1) and the oscillating bubbles (life phase 2) precede the usually observed plasmonic bubble dynamics \cite{wang2017, baffou2014jpc, liu2015, adleman2009, chen2017}. As revealed in \cite{wang2017}, this later plasmonic bubble dynamics consists of two subsequent and slower phases, namely a  vaporization-dominated growth (life phase 3) and a diffusion-driven growth that is dominated by the influx of gas dissolved from the water (life phase 4), and correspondingly depends on the dissolved gas concentration. Fig. \ref{fig:table} summarizes all four life phases.

%\vspace{-40pt}

\begin{figure}[h]
	\centering
	\includegraphics[width=0.4\columnwidth]{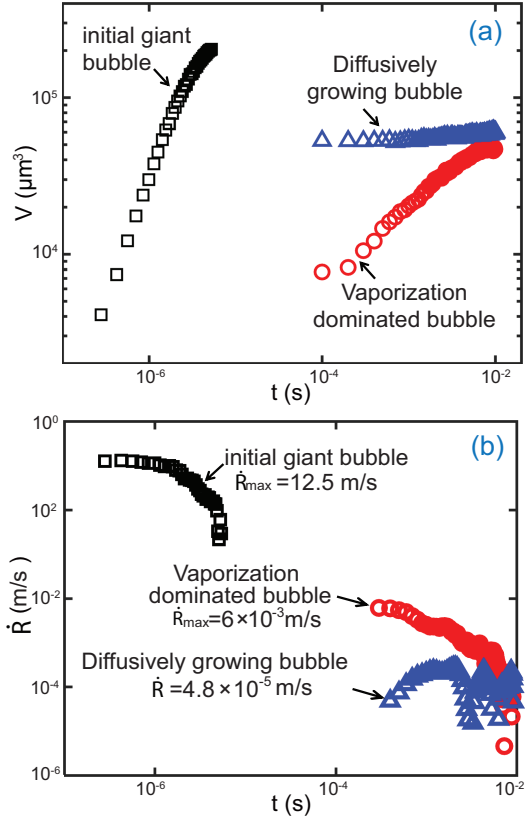}
	\caption{
		(a) Bubble volume dynamics during growth for an initial giant bubble (life phase 1, black), a vaporization dominated bubble (life phase 3, red), and a diffusively growing bubble (life phase 4, blue) in gas-rich water. (b) Growth rates $\dot{R}$ of the same three bubbles. Both plots are in double logarithmic scale. Note: the origin of time for the diffusively growing bubble was aligned to that of vaporization dominated bubble to facilitate the comparison.
	}
	\label{fig:growth rates}
\end{figure}

Fig. \ref{fig:growth rates}a shows the volume $V$ as a function of time $t$ for an initial giant bubble (life phase 1), a vaporization dominated bubble (life phase 3), and a bubble growing slowly by gas diffusion (life phase 4) for the gas-rich case. The volume of the initial giant bubble rapidly exceeds that of the vaporization dominated bubble (phase 3) and that of the diffusively growing bubbles (phase 4). The growth rate of the giant bubble (see Fig. \ref{fig:growth rates}b) reaches a maximum value of about 12.5 m/s, which is respectively about 2000 times and $10^5$ times larger than the respective growth rates of the vaporization dominated and the diffusively growing bubbles.

\begin{figure*}[h]
% 		\vspace{-30pt}
 	\includegraphics[width=1\textwidth]{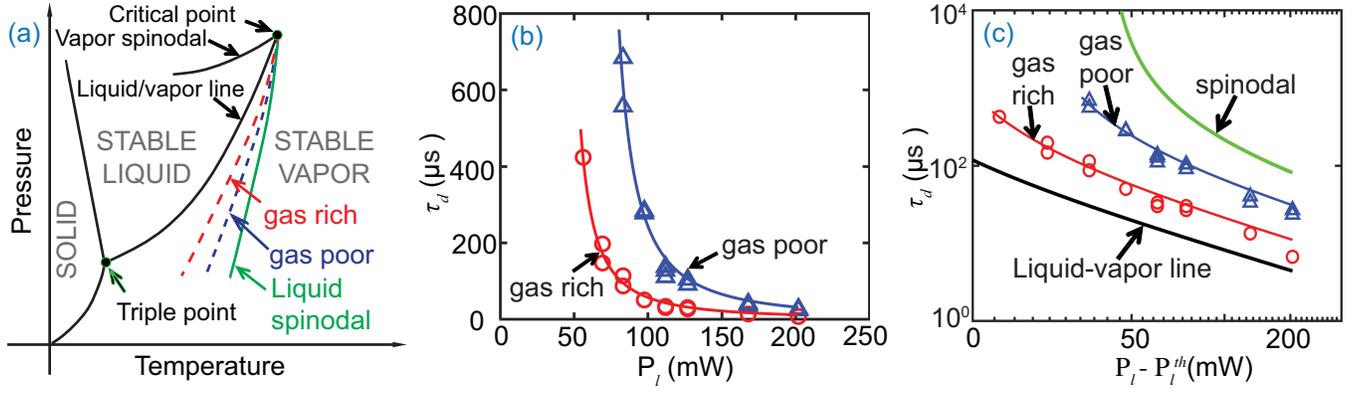}
 	\caption{
 		(a) Phase diagram of water (schematics). The green solid line is the liquid spinodal line, the theoretical limit of superheat, while the blue and red dashed lines schematically depict the attainable superheat for gas-poor and gas-rich water. (b) Measured delay $\tau_d$ as function of $P_l$. The symbols represent the experimental data and the solid lines the fit curves using Eq.(\ref{heat_conduction}). (c) Double logarithmic plot of $\tau_d$ vs $P_l-P_{th}$. Both curves fall within the theoretical limits, namely the boiling temperature (black curve) and the spinodal curve ($T_s$ = 578.2 K for a pressure of 1 atm, green curve). The shorter delay time $\tau_d$ for gas-rich water indicates that dissolved gas facilitates bubble nucleation.
 	}
 	\label{fig:delay}
% 	\vspace{-30pt}
 \end{figure*}

In order to capture the dynamics on a longer timescale that encompasses the nucleation delay, a second set of experiments was performed using a high speed camera operated at 300 kfps (Supplemental material). The experiments were conducted with different laser powers $P_l$ in both gas-rich water and gas-poor water. The observed decrease in maximum volume $V_{max}$ of the initial giant bubble with increasing laser power $P_l$ seems counter-intuitive and so does the observed larger bubble volume for the gas-poor case. The reason for this behavior is that nucleation of vapor bubbles in water requires the temperature to reach the nucleation temperature $T_n$. Under ideal conditions (pure water), this temperature is identical to the liquid spinodal decomposition temperature \cite{skripov1970}. However, the presence of impurities, gases or interfaces, results in a lower $T_n$ (see Fig.\ref{fig:delay}a). The nucleation temperature $T_n$ does not depend on the laser power $P_l$, which explains the increased delay time $\tau_d$ for lower $P_l$ seen in Fig. \ref{fig:delay}b.

 %\vspace{-20pt}

We now go ahead and analytically quantify this behavior. The time-dependent temperature field $T(\vec{r}, t)$ around a single nanoparticle, assuming spherical geometry and constant thermal properties, is governed by the spherical linear Fourier equation for heat conduction:
\begin{equation}
	\partial_t(T(r,t)) = \frac{p_l(r,t)}{\rho c_p} + \kappa\frac{1}{r^2}\partial_r(r^2\partial_rT(r,t)),
	\label{heat_conduction}
\end{equation}
\noindent where $\kappa$, $\rho$, and $c_p$ are thermal diffusivity, density, and heat capacity of water, $r$ is the spherical distance to the GNP, and $p_l(r,t)$ is the deposited power density (unit in $W/m^3$), which is assumed to be constant for a radius $r$ within the GNP, and 0 elsewhere.

%\textcolor{red}{We now go ahead and analytically quantify this behavior. The time-dependent temperature field $T(\vec{r}, t)$ around a single nanoparticle, assuming spherical geometry and constant thermal properties, is governed by the spherical linear Fourier equation for heat conduction:
%\begin{equation}
%	\partial_t(T(r,t)) = \frac{p_l}{\rho c_p} + \kappa\frac{1}{r^2}\partial_r(r^2\partial_rT(r,t)).
%	\label{heat_conduction}
%\end{equation}
%\noindent Here $\kappa$, $\rho$, and $c_p$ are thermal diffusivity, density, and heat capacity of water, $r$ is the spherical distance to the GNP, and $p_l$ is the deposited power density (unit in $W/m^3$) on the GNP. Since GNPs are two orders of magnitude smaller than the laser beam, for a specific GNP, the power density on it can be taken as a constant value.}

This problem is solved analytically, in the Fourier domain, and subsequently reversed back to real space numerically. The temperature field generated by the nanoparticle array can then be computed by superposition, placing the nanoparticle sources on the liquid/substrate interface, within the Gaussian laser beam. A first order correction is applied to account for the presence of a substrate as detailed in the Supplementary Material.

The resulting time-dependent temperature field is the linear superposition of the temperature fields of the $N_{np}$ nanoparticles,
\begin{equation}
	T(x,y,z,t) = \sum_{n=1}^{N_{np}}\left(T_i(d_{i,(x,y,z)},t)\right),
\end{equation}
with $T_i$ the temperature field created by the particle i, $d_{i,(x,y,z)}$ the distance from the center of this nanoparticle to the point located at the coordinates $(x,y,z)$.
The result, proportional to the input power, is given in the Supplementary Material. From the computation, taking into account the laser input power, one directly obtains the time required to reach a given temperature for a given laser power. This approach was used to fit the experimental data in Fig. \ref{fig:delay}b using a root-mean-square-minimization method, resulting in the solid curves in Fig. \ref{fig:delay}b.

This fitting procedure directly provides values for the nucleation temperature, namely $T_n = 422$ K and $T_n = 498$ K for the gas-rich and gas-poor water, respectively, and for the vaporization power thresholds $P_l^{th}$, namely $P_l^{th}$ = 39 mW and 62 mW, respectively. Figure \ref{fig:delay}c shows a double-logarithmic plot of $\tau_d$ versus $P_l-P_l^{th}$. As expected, both curves are located between the two limiting cases, namely the liquid-vapor equilibrium temperature $T_n$ = 373.2 K (black curve) and the water spinodal temperature $T_s$ = 578.2 K \cite{skripov1970} (green curve). Moreover, above obtained values for the vaporization power thresholds are in reasonable agreement with the respective measured thresholds of 44 mW and 56 mW for the gas-rich and gas-poor case. Below this threshold, the steady-state regime for spherical heat diffusion has time to establish and the temperature stops rising before the system reaches the required nucleation temperature.

\begin{figure*}[b]
%	\vspace{-20pt}
	\centering
	\includegraphics[width=0.8\textwidth]{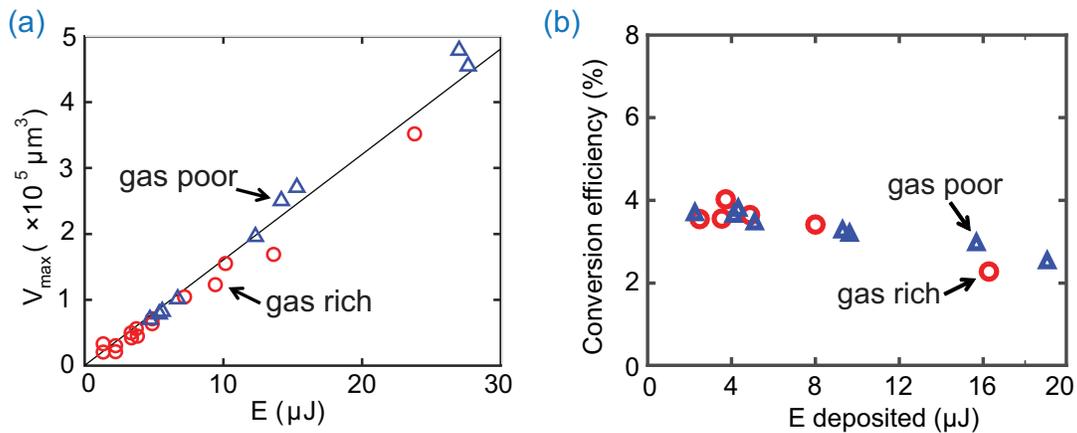}
	\caption{
		(a) Maximal volume of the giant bubble $V_{max}$ as function of the energy $E = P_l \tau_d$ in gas-rich and gas-poor water. Both cases show an identical linear relation between $V_{max} =  kE$, regardless of $\tau_d$ and $P_l$. (b) Actual conversion efficiency of the GNPs for the energy converted from laser heat deposition to vaporization enthalpy contained in the vapor.
	}
	\label{fig:volume and energy}
	\vspace{-10pt}
\end{figure*}
%\vspace{-20pt}

The experimental results also reveal that the dissolved gas plays a crucial role in the initial giant bubble nucleation. Numerous studies have shown that impurities in water can greatly reduce the nucleation temperature $T_n$ from the liquid spinodal temperature \cite{skripov1974, blander1975, puchinskis2001}. As the concentration of dissolved gas in the gas-poor water is about half of that in gas-rich water, the probability of forming gas nuclei larger than the critical size is statistically reduced \cite{caupin2006}, resulting in a higher $T_n$, which leads to an increase in the delay time $\tau_d$ in gas-poor water.

Nonetheless, as shown in Fig. \ref{fig:volume and energy}a, the maximum bubble volume $V_{max}$ displays a universal linear relation $V_{max} = kE$ (with $k\approx $ $1.7\times$ $10^4$ $\upmu$m$^3/\upmu$J) with the total dumped energy $ E = P_\ell \tau_d$, which is the accumulated laser energy in the illumination spot on the substrate from the moment the laser was switched on to the moment of bubble nucleation. The very same linear relation holds for both gas concentrations. This reflects that the energy stored in the vicinity of the nucleus determines the energy available for vaporization, and more energy results in larger vapor bubbles. The linear relation is a consequence of the short delay time observed in these experiments relatively to the thermal diffusion time $\tau_{diff} \approx R_l^2/\pi/\kappa \approx 400 \upmu$s, where $R_l$ is the laser spot radius. Moreover, it further confirms that the initial giant bubbles are pure vapor bubbles, for both gas-poor and gas-rich water.

The major motivation for using plasmonic particles for applications such as solar to steam energy harvesting or plasmonic bubble photoacoustic therapy lies in their outstanding efficiency of light absorption. In such case, the limiting factor becomes the thermal processes occurring within the system, that convert thermal energy into vapor. It is therefore important to quantify the energy conversion efficiency.
For water, one can neglect the heat capacity compared to latent heat of vaporization. Thus, the energy contained in the giant initial bubble can be estimated using the ideal gas law:
\begin{equation}
	E_{bub} = \Lambda_{vap} \dfrac{M P_{sat} V_{max}}{R_g T_{sat}},
\end{equation}
where $P_{sat}$ and $T_{sat}$ are water saturation pressure and temperature, respectively, $M$ is the molar mass of water, $\Lambda_{vap}$ its latent heat of vaporization and $R_g$ = 8.314 $J mol^{-1}K^{-1}$ the gas constant. Since the ratio $P_{sat}/T_{sat}$ is independent of the laser power $P_l$ (see Supplementary Material), $E_{bub}$ is proportional to the maximal bubble volume $V_{max}$. The ratio of the energy contained in the initial giant vapor bubble to the energy absorbed by the substrate, which is the effective energy conversion efficiency for this process, is displayed in Fig.\ref{fig:volume and energy}b and is equal to ($3.6 \pm 0.5$) $\%$. The conversion efficiency displays a slight decrease for larger energies, corresponding to an increase of $\tau_d$. This is in agreement with the increasing (with time) losses by heat diffusion.

To summarize, we have shown that the nucleation of plasmonic bubble on water-immersed, laser-irradiated GNPs is initiated by a transient and explosively growing giant vapor bubble with a life time of about 10 $\upmu$s. The maximum growth rate $\dot{R}$ of the initial giant bubbles exceeds 12.5 $m/s$, which is three orders of magnitude larger than that of the later and hitherto observed plasmonic bubbles that grow by steady vaporization. Whether and when a giant initial bubble nucleates is determined by the competition between laser heating and cooling through thermal diffusion. As a result, the delay time $\tau_d$ up to bubble nucleation decreases with increasing laser power $P_l$, leading to smaller bubbles. Both the nucleation temperature $T_n$ and the laser power threshold value can be obtained from a simple heat diffusion model, and are consistent with the experimental values. Moreover, the experimental results show that the gas-poor water has a much larger delay time $\tau_d$ than the gas-rich water. This reflects that the dissolved gas facilitates vapor bubble nucleation and lowers the superheat temperature limit. After nucleation, the giant bubbles in both cases obey the same dynamics (life phase 1). The maximum bubble volume follows the same linear relation with the accumulated energy for both gas-rich and gas-poor water. In the later life stages, the plasmonic bubble displays small, sustained oscillations (life phase 2), followed by the known vaporization dominated phase (life phase 3), and diffusive growth phase (life phase 4). Our findings on plasmonic bubble dynamics have strong bearings on various applications of plasmonic bubbles, notably on medical applications where large plasmonic bubbles can cause damage \cite{brennen1995}. In the context of catalysis or triggering chemical reactions the energetic giant bubble collapse may be beneficial.

\matmethods{
\subsection*{Sample Preparation}
A gold layer of approximately 45 nm was deposited on an amorphous fused-silica wafer by using an ion-beam sputtering system (home-built T$^\prime$COathy machine, MESA+ NanoLab, Twente University). A bottom anti-reflection coating (BARC) layer ($\sim$186 nm) and a photoresist (PR) layer ($\sim$200 nm) were subsequently coated on the wafer. Periodic nanocolumns with diameters of approximately 110 nm were patterned in the PR layer using displacement Talbot lithography (PhableR 100C, EULITHA) \cite{the2017}. These periodic PR nanocolumns were subsequently  transferred at wafer level to the underlying BARC layer, forming 110 nm BARC nanocolumns by using nitrogen plasma etching (home-built TEtske machine, NanoLab) at 10 mTorr and 25 W for 8 min. Using these BARC nanocolumns as a mask, the Au layer was subsequently etched by ion beam etching (Oxford i300, Oxford Instruments, United Kingdom) with 5 sccm Ar and 50-55 mA at an inclined angle of $5^{\circ}$. The etching for 9 min resulted in periodic Au nanodots supported on cone-shaped fused-silica features. The remaining BARC was stripped using oxygen plasma for 10 min (TePla 300E, PVA TePla AG, Germany). The fabricated array of Au nanodots was heated to $1100^{\circ}$C in 90 min and subsequently cooled passively to room temperature. During the annealing process, these Au nanodots re-formed into spherical-shaped Au nanoparticles.

\begin{figure}[h]
	\centering
	\includegraphics[width=0.7\textwidth]{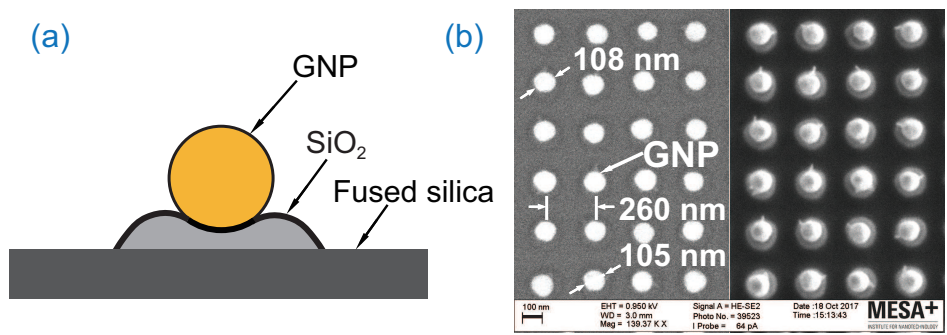}
	\caption{	
		(a) Schematic of a gold nanoparticle sitting on a $SiO_2$ island on a fused-silica substrate. (b) SEM images of the patterned gold nanoparticle sample surface. Left: Energy Selective Backscatter (ESB) mode. Right:  High Efficiency Secondary Electron mode.
	}
	\label{fig:gnp}
\end{figure}

\subsection*{Setup description}
In this study, two imaging setups were used to capture the growth dynamics of plasmonic microbubbles, both on short-term and long-term, as shown in Fig. \ref{fig:setups}. Short-term measurements were performed using the Brandaris 128 ultra-fast imaging system \cite{chien2003, gelderblom2012}. This system can capture 128 consecutive images with a frame rate of up to 25 Mfps. The schematics of the Brandaris 128 setup is shown in Fig. \ref{fig:setups}(a); an upright microscope was installed together with a water immersion objective (LUMPLFLN, Olympus) for bubble observation. In the Brandaris 128, 128 CCD cameras are sequentially installed along an arc. The images from the objective are redirected to the sequence of the CCD sensors by a rotating mirror-polished beryllium turbine. By adjusting the rotation speed of the turbine one can tune the recording speed. In our experiments the frame rates were around 7-8 MHz, which allowed to capture the detailed temporal evolution of the initial giant bubbles or 3 to 4oscillation cycles of the subsequent oscillating bubbles. A xenon flash light was used as illumination source.

\begin{figure*}[t]
	\centering
	\includegraphics[width=0.8\textwidth]{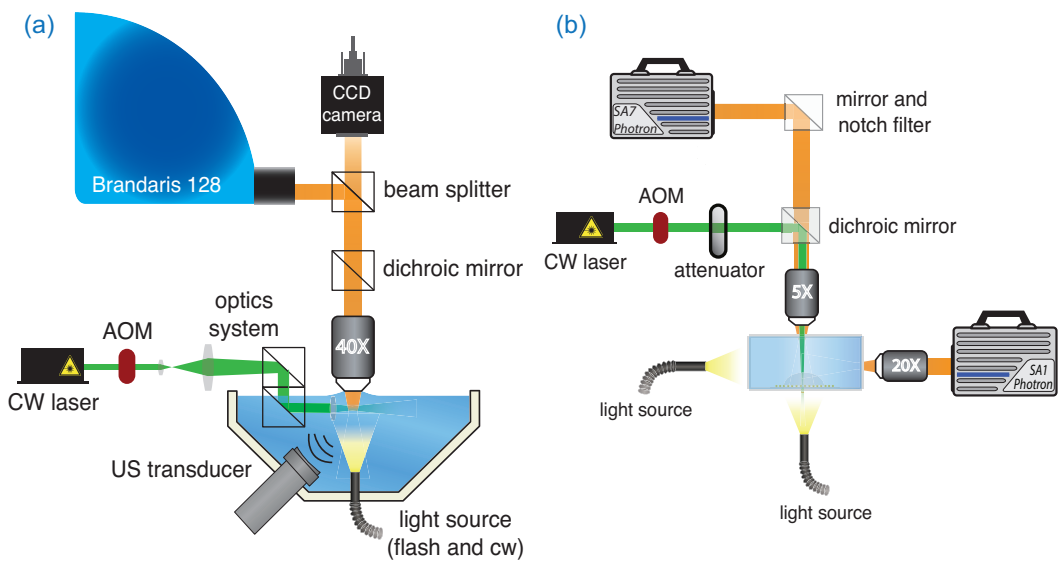}
	\caption{	
		Schematic of the optical imaging facilities for plasmonic microbubble formation observation. (a) Brandaris 128 imaging system with frame rate up to 25 Mfps. (b) High-speed camera imaging system with frame rate up to 500 kfps.
	}
	\label{fig:setups}
\end{figure*}

The gold nanoparticle decorated sample was immersed in a water tank and placed vertically to enable side view imaging of plasmonic bubbles. During measurement, the sample surface was irradiated with a continuous-wave laser (Cobolt Samba) of 532 nm wavelength and tuneable power up to 200 mW. An acousto-optic modulator (Opto-Electronic, AOTFncVIS) was used as a shutter to control on/off of laser irradiation on the sample surface. A 400 $\upmu$s laser pulse was generated and controlled by pulse/delay generator (BNC model 565). Additionally, an ultrasonic transducer was installed inside the water tank in order to obtain the acoustic signal of initial bubble nucleation. This facilitates the synchronization of the laser pulse and Brandaris 128 image capture.

The Brandaris 128 ultra-high-speed imaging facility can capture 128 consecutive frames, thus recording only for a limited period of time. In order to capture the dynamics on a larger time scale, a second setup was designed with a high-speed camera operated at a frame rate of 300 Kfps, as depicted in Fig. \ref{fig:setups}b, equipped with 5x (LMPLFLN, Olympus) and 20x (SLMPLN, Olympus) long working distance objectives. The 5X objective was used to focus laser onto the sample surface, the other one was used to obtain side-view images. Bubble growth was recorded using a high-speed camera (Photron SA1) operated at framerate of up to 500 Kfps. Another high-speed camera (Photron SA7) was used to capture top view images of the sample surface for optical alignment. Two light sources, Olympus ILP-1 and Schott ACE I, were used to provide illumination for both high-speed cameras.

In two sets of experiments, plasmonic microbubble formation in both gas-rich water and gas-poor water was studied. Water directly obtained from a Milli-Q machine was taken as gas-rich water. To get gas-poor water, the liquid cell filled with water and vacuumed. The total degassing time was about 3 hours. The relative gas concentration for both gas-rich and gas-poor water was measured with an oxygen meter (Fibox 3 Trace, PreSens) in the ambient environment (temperature: 22$^{\circ}$C). During measurements, the fluid cell was sealed to slow down re-gassing. The relative gas concentration in the gas-poor water is 34\%. This is about a half of the value 65\% measured in the gas-rich water.

}

\showmatmethods{}

\acknow{We thank Andrea Prosperetti for helpful discussions. The authors also thank the Dutch Organization for Research (NWO) and the Netherlands Center for Multiscale Catalytic Energy Conversion (MCEC) for financial support. Y.W. appreciates financial support from National Natural Science Foundation of China (Grant No. 51775028) and Beijing Natural Science Foundation (Grant No. 3182022).}

\showacknow{} % Display the acknowledgments section

% \pnasbreak splits and balances the columns before the references.
% Uncomment \pnasbreak to view the references in the PNAS-style
% If you see unexpected formatting errors, try commenting out \pnasbreak
% as it can run into problems with floats and footnotes on the final page.
%\pnasbreak

% Bibliography
\bibliography{biblioPNAS}

\end{document}